\documentclass[11pt]{article}

\usepackage{amsmath,amssymb,amsthm,amstext}

\usepackage{graphicx,epsfig}
\usepackage{fullpage}

\renewcommand{\le}{\leqslant}
\renewcommand{\ge}{\geqslant}

\newtheorem{theorem}{Theorem}
\newtheorem{lemma}{Lemma}

\theoremstyle{remark}
\newtheorem{defi}{Definition}

\newtheorem{con}{Conjecture}
\newtheorem{que}{Question}

\newcommand{\N}{\mathbb N}

\newcommand{\MA}{\text{MA}}
\newcommand{\poly}{\text{poly}}
\newcommand{\heurMA}{\text{Heur-MA}}
\newcommand{\heurAM}{\text{Heur-AM}}

\newcommand{\heurPPNP}{\text{Heur-PP}\!\cdot\!\text{NP}}
\newcommand{\AM}{\text{AM}}
\newcommand{\MAP}{\text{MAP}}
\newcommand{\PP}{\text{PP}}
\newcommand{\BPP}{\text{BPP}}
\newcommand{\NP}{\text{NP}}
\newcommand{\PPNP}{\text{PP}\!\cdot\!\text{NP}}
\newcommand{\p}{\text{P}}
\newcommand{\prob}{\text{Prob}}

\title{How Much Randomness is Needed to Convert MA Protocols to AM Protocols?}

\author{Nikolay Vereshchagin\thanks{The results are supported by Russian Science Foundation (20-11-20203).}\\
Moscow State University, HSE University, Russian Federation, and Yandex.}
\date{}

\begin{document}

\maketitle
\begin{abstract}
The Merlin-Arthur  class of languages MA is included into Arthur-Merlin  class AM, and into PP.
For a  standard transformation of a given MA protocol with 
Arthur's message (= random string) of length $a$ and Merlin's message of length $m$ to a PP machine,
the latter needs  $O(ma)$ random bits. The same holds for simulating
MA protocols by AM protocols: in the resulting AM protocol the length of Arthur's message (= random string) is 
$O(ma)$. And the same holds for simulating heuristic 
MA protocols by  heuristic  AM protocols as well. In the paper  [A. Knop, Circuit Lower Bounds for Average-Case MA, 
CSR 2015] it was conjectured that, in the transformation of heuristic MA protocols to heuristic AM protocols,
 $O(ma)$ can be replaced by a polynomial of $a$ only.
A similar question can be asked for normal MA and AM protocols, and for the simulation of MA protocols by PP machines.
In the present paper we show that, relative to an oracle,  both latter questions answer in the negative and Knop's conjecture is false.
Moreover,  the same is true for simulation of MA protocols by AM protocols in which the  error probability is not bounded away from 1/2,
the so called PP$\cdot$NP protocols. The latter protocols generalize both AM protocols and  PP machines.
  \end{abstract}  
 
\section{Introduction}

Let $\MA[m,a]$ denote the class of languages that have two-round 
Merlin-Arthur  protocols in  which  Merlin starts the communication,
the lengths of messages are $m,a$, respectively, and error probability  is at most $1/3$.
In a similar way the class $\AM[m,a]$ is defined, but  now  Arthur is the one starting the communication.

The class $\AM$ includes $\MA$ \cite{babai85},
more specifically, it holds that  $\MA[m,a]\subset \AM[m,O(ma)]$.
This can be proved via amplification: we first decrease the error probability  
from 1/3 to $2^{-m-1}$, at the expense of increasing
 $a$ to $O(ma)$, as Arthur has to repeat his algorithm 
$O(m)$ times and then make a majority vote.
Using derandomization via expanders \cite{alon1995derandomized}, we can
replace $O(ma)$ by $a+O(m)$ and hence prove the inclusion $\MA[m,a]\subset \AM[m,a+O(m)]$.
In this paper we try to understand whether it is possible get rid of  $O(m)$ and replace in this inclusion $a+O(m)$ by a polynomial of 
$a$ only. 

\begin{que}\label{q1}
Is there a polynomial $p$ such that 
$\MA[m,a]\subset \AM[*,p(a)]$ for all polynomials $m,a$ of the length of the input?
\end{que}
Here $\AM[*,a]$ denotes the union of $\AM[m,a]$ over all polynomials $m$.
This question is motivated by the following conjecture from
Knop's paper~\cite{knop2015} about heuristic analogs of MA and AM.
\begin{con}[\cite{knop2015}]\label{con}
There is a polynomial $p$ such that if there is a heuristic Merlin-Arthur protocol for a language $L$
that on inputs of length $n$ and confidence $\delta$ uses $q(n/\delta)$ random bits
($q$ is a polynomial) then there is a heuristic  Arthur-Merlin protocol for 
$L$ using $p(q(n/\delta))$ random bits.
\end{con}
Roughly speaking, a language $L$ has a heuristic Merlin-Arthur protocol 
if there is an ordinary  Merlin-Arthur protocol that errs only on a small fraction of inputs (called its confidence).
The heuristic analog of AM is defined in a similar way.

This conjecture is interesting, since it provides a way to 
transform lower bounds for heuristic computations  to lower bounds for
normal computations. More specifically, the following holds:
if Conjecture~\ref{con} is true then for all $k$ some NP language has no Boolean circuits of size $n^k$ (\cite{knop2015}).  
Thus it would be very helpful to prove Conjecture 1.


A similar question arises for the inclusion $\MA\subset\PP$ \cite{verPP},
where $\PP$ stands for the class of languages recognized by probabilistic polynomial time machines with 
error probability  less than 1/2. It is important that the error probability  is not bounded away from 1/2.
This inclusion is also proved via amplification and  its more detailed version reads
$\MA[m,a]\subset \PP[O(ma)]$, where $r$ in the notation $\PP[r]$ denotes the number of random bits available to PP machines.
\begin{que}\label{q2}
Is there a polynomial $p$ such that 
$\MA[m,a]\subset \PP[p(a)]$ for all polynomials $m,a$?
\end{que}

If $\MA=\p$ or $\MA=\NP$, 
then in both above simulations randomness is not needed at all, 
therefore both questions answer in positive. Hence to prove negative answers, we must show beforehand that $\MA\ne\p$. The latter is equivalent to $\p\ne\NP$ since $\MA\subset\Pi_2^p$ \cite{BHZ}.
Thus we cannot hope to prove negative answers to both Questions 1 and 2 unless we show that  $\p\ne\NP$. 
On the other hand, positive answers seem implausible.

In such a situation, it is natural to ask whether 
one can answer Questions 1 and 2 using relativizable techniques.
By the result of~\cite{BGS} there is an oracle under which $\p\ne\NP$. Under that oracle
$\MA=\p$, since the Polynomial Hierarchy collapses. Hence under that oracle
both questions answer in the positive and Knop's conjecture holds.

On the other hand, in the present paper we show that there are oracles 
under which both questions answer in the \emph{negative} and Knop's conjecture does \emph{not} hold.
More specifically, we show that 
there is an oracle  under which for every polynomial $p$ the class $\MA[m,a]$ is not included 
in both classes $\AM[*,p(a)]$ and $\PP[p(a)]$ where $a(n)=n$, $m(n)=p(n)$, and  
$n$ stands for the length of the input.
In particular, under that oracle $\MA\ne \NP$ thus the ``full derandomization" of $\MA$ is impossible.  Our result implies that  
we need a non-relativizable proof techniques to fully derandomize $\MA$ and even to show that $\MA[m,a]\subset \AM[*,p(a)]$ for some fixed polynomial $p$. 
It remains an open question whether we can resolve both questions using algebrizable techniques in the sense of Aaronson and Wigderson~\cite{AW}.  
 
Actually, we prove a stronger separation. 
Let $\PPNP[m,a]$ denote the class of languages recognized by Arthur-Merlin   protocols with  
error probability  less than 1/2 but not bounded away from 1/2. 
This class obviously includes both classes
$\PP[a]$ and 
$\AM[m,a]$.
%
%
In the present paper we prove that under an oracle,
for all non-constant polynomials $m,a$ there is a language $L$ in $\MA[m,a]$ such that for every $\PPNP [m',a']$ protocol for $L$ it holds  
$a'(n)\ge m(n)+a(n)-O(\log n)$ for  all $n$. 

Using the same techniques, we then show that under an oracle Knop's conjecture is false, too.
We also establish a similar theorem for MA protocols,
where Merlin never fails. More specifically,
let  $\MAP[m,a]$ denote the subclass of $\MA[m,a]$ consisting of languages possessing a protocol for which the error probability is zero for all 
strings from the language. 
It is natural to ask whether  $\MAP[m,a]\subset \AM[*,\poly(a)]$. 
From the Sipser--G\'acs--Lautemann theorem~\cite{sipser-bpp,lauteman} about the inclusion $\BPP\subset\Pi_2$ it follows that $\MAP=\MA$,
more specifically, 
 $$
 \MA[m,a]\subset \MAP[O((m+a\log a)a\log a),O((a\log a)^2)].
 $$ 
 Hence our theorem implies that under an oracle  for all polynomials $p$ there are polynomials $m,a$
with $\MAP[m,a]\not\subset \AM[*,p(a)]$.
In the present paper, we establish a more tight bound:   relative to an oracle, 
for every non-constant polynomials $m,a$ there is a language $L$ in $\MAP[m,a]$ such that 
for any $\PPNP [*,a']$ protocol for $L$ we have  
$a'(n)\ge \max\{m(n),a(n)\}-O(\log n)$ for almost all $n$.
Compared with simulating  $\MA$ protocols, $m(n)+a(n)$ is replaced by
 $\max\{m(n),a(n)\}$.

Summarizing, we show that we cannot prove Conjecture 1 or positive answers to Questions 1 and 2 using relativizable techniques.

\section{Preliminaries}

\emph{A language} is a subset of the set $\{0,1\}^*$ of all strings over the binary alphabet. The length of the string
$x$ is denoted by 
 $|x|$. 
 When we speak on probability, we always mean the uniform distribution.
 
A \emph{polynomial} is a function $p:\N\to\N$ of the form $p(n)= b(n+1)^c$ where $b,c$ are positive integers.
Thus any polynomial is a non-constant function with positive values.
In the following definitions $r,m,n,t$ denote some polynomials.

\begin{defi}
A language  $L$ is in the class  $\PP[r]$ if there is a Turing machine
$V$ with input strings $x,a$, whose running time is bounded by some polynomial of
$|x|$, such that 
$$
x\in L\leftrightarrow \prob_{a\in\{0,1\}^{r(|x|)}}[V(x,a)=1]>1/2
$$
for all strings $x$.
\end{defi}

\begin{defi}
A language  $L$ is in the class  
$\MA[m,a]$ if  there is a Turing machine
$V$ with input strings $x,y,z$, whose running time is bounded by some polynomial of
$|x|$, such that  
\begin{align*}
x\in L&\rightarrow \exists y\in\{0,1\}^{m(|x|)}\prob_{z\in\{0,1\}^{a(|x|)}}[V(x,y,z)=1]>2/3,\\
x\notin L&\rightarrow \forall y\in\{0,1\}^{m(|x|)}\prob_{z\in\{0,1\}^{a(|x|)}}[V(x,y,z)=1]<1/3.
\end{align*}
for all strings $x$.
\end{defi}

\begin{defi}
A language  $L$ is in the class  
$\MAP[m,a]$ if  there is a Turing machine
$V$ with input strings $x,y,z$, whose running time is bounded by some polynomial of
$|x|$, such that  
\begin{align*}
x\in L&\rightarrow \exists y\in\{0,1\}^{m(|x|)}\prob_{z\in\{0,1\}^{a(|x|)}}[V(x,y,z)=1]=1,\\
x\notin L&\rightarrow \forall y\in\{0,1\}^{m(|x|)}\prob_{z\in\{0,1\}^{a(|x|)}}[V(x,y,z)=1]<1/2.
\end{align*}
for all strings  $x$.
Compared with the definition of $\MA$, the number  2/3 is replaced by  1,
and the number 1/3 by 1/2 (the latter replacement is not important).
\end{defi}

\begin{defi}
A language  $L$ is in the class  
$\AM[m,a]$ if  there is a Turing machine
$V$ with input strings $x,y,z$, whose running time is bounded by some polynomial of
$|x|$, such that 
\begin{align*}
x\in L&\rightarrow \prob_{z\in\{0,1\}^{a(|x|)}}[\exists y\in\{0,1\}^{m(|x|)} V(x,y,z)=1]>2/3,\\
x\notin L&\rightarrow \prob_{z\in\{0,1\}^{a(|x|)}}[\exists y\in\{0,1\}^{m(|x|)} V(x,y,z)=1]<1/3.
\end{align*}
for all strings $x$.
\end{defi}

\begin{defi}
A language  $L$ is in the class  
$\PPNP[m,a,t]$  if  there is a Turing machine
$V$ with input strings $x,y,z$, whose running time is bounded by 
$t(|x|)$, such that 
\begin{align}\label{eqm}
x\in L&\leftrightarrow \prob_{z\in\{0,1\}^{a(|x|)}}[\exists y\in\{0,1\}^{m(|x|)} V(x,y,z)=1]>1/2.
\end{align}
for all strings $x$.
Triples of the form $(m,a,V)$ will be called \emph{$\PPNP$ protocols}. 
We say that a $\PPNP$ protocol \emph{is correct  on input $x$} if
the equivalence~\eqref{eqm} is true. Otherwise we say that the protocol \emph{errs on $x$}.
\end{defi}

\begin{defi}(Heuristic classes)
We say that a  language $L$ is in $\heurPPNP[m,a]$ 
if  there is a Turing machine
$V$ with input strings $x,y,z,\delta$, whose running time is bounded by 
a polynomial of $|x|/\delta$, such that for all $n$ the equivalence
\begin{align}\label{eq0}
x\in L&\leftrightarrow \prob_{z\in\{0,1\}^{a(|x|/\delta)}}[\exists y\in\{0,1\}^{m(|x|/\delta)} V(x,y,z,\delta)=1]>1/2.
\end{align}
holds for all  but $\delta$ fraction of strings $x$ of length $n$.

In a similar way the classes $\heurMA[m,a]$ and  $\heurAM[m,a]$ are defined. 
\end{defi}

\emph{An oracle} is a function $A:\{0,1\}^*\to\{0,1\}$. The classes
 $\MA,\MAP,\AM,\PPNP$, \emph{relativized by an oracle $A$} are defined as follows: now the 
 machine $V$ has an extra  ``oracle'' tape. On that tape the machine can
 ``query the oracle''. That means that the machine can write any string $u$ followed by a question mark on that
tape. Immediately after that, the word $u$ is replaced by $A(u)$ by ``the oracle'', which is counted as one step of computation.
Complexity classes relativized by an oracle  $A$ are denoted by
$\MA^A,\MAP^A,\AM^A,\PPNP^A$. 

\section{Constructing an Oracle under which Transformation of $\MA$ to $\PPNP$
Protocols Requires Many Random Bits}

The first result states that 
under some oracle some language in $\MA[m,a]$ has no  
$\PPNP[m',a',t']$ protocols unless  
$a'(n)\ge m(n)+a(n)-\log_2 t'(n)-O(1)$.  

\begin{theorem}\label{th1}
There is an oracle $A:\{0,1\}^*\to\{0,1\}$ with the following property.
For all polynomials 
$m(n),a(n)$ 
there is a language in $\MA^{A}[m,a]$ that is outside any 
class $\PPNP^A[m',a',t']$ such that 
$a'(n)< m(n)+a(n)-\log_2(3t'(n))$ for infinitely many $n$. 
\end{theorem}
\begin{proof}
We first prove a weaker version of the theorem, assuming that the polynomials 
$m(n),a(n)$ are fixed.

The value of an oracle $A$
on strings of length $m(n)+a(n)$ can be viewed as a Boolean matrix 
$A_n$ with   
 $2^{m(n)}$ rows and 
 $2^{a(n)}$ columns. For the constructed oracle 
 $A$ for all $n$ the matrix $A_n$ will have one of the following forms:
\begin{itemize}
\item either some row in the matrix  has more than two thirds of ones, 
\item or less than  one third of entries in every row are  ones.
\end{itemize}
Matrices of the first type are called \emph{heavy},
and matrices of the second  type are called \emph{light}.
\begin{figure}
\begin{center}
$$
\begin{pmatrix}
0&0&1&0&0&0&0\\
0&0&0&0&0&0&0\\
0&0&0&0&1&0&0\\
0&0&0&0&0&0&0\\
1&0&1&1&0&1&1\\
0&0&0&0&0&0&0
\end{pmatrix} 
\qquad\qquad\qquad\qquad
\begin{pmatrix}
0&1&1&0&0&0&0\\
1&0&1&0&0&0&0\\
0&0&0&0&0&0&1\\
0&0&0&1&0&0&1\\
0&0&0&0&0&0&0\\
0&0&0&0&0&1&1
\end{pmatrix} 
$$
\end{center}
\caption{The matrix on the left is heavy, and  the matrix on the right is light.}\label{t}
\end{figure}
We will call this property of $A$ by $P_{m,a}$ (see Fig.~\ref t).

The language $L=L(A)$ will consist of all strings
$1^n$ such that the matrix 
$A_n$ is heavy. The intuition is the following: the hidden heavy row is easy to guess in one step but hard to amplify and find.

The property $P_{m,a}$ of $A$ guarantees that 
$L(A)\in\MA^{A}[m,a]$, since
\begin{align*}
1^n&\in L(A)\rightarrow \exists y\in\{0,1\}^{m(n)}\prob_{z\in\{0,1\}^{a(n)}}[A(yz)=1]>2/3,\\
1^n&\notin L(A)\rightarrow \forall y\in\{0,1\}^{m(n)}\prob_{z\in\{0,1\}^{a(n)}}[A(yz)=1]<1/3
\end{align*}
(the respective machine $V$ on input 
$(1^n,y,z)$ outputs $A(yz)$ by querying the oracle once).

Now we will define an oracle $A$ so that $L(A)$ is outside 
$\PPNP^A[m',a',t']$ provided 
$a'(n)< m(n)+a(n)-\log3 t'(n)$
for infinitely many $n$. 

We first pick any $\PPNP$ protocol  $(m',a',V)$ with 
$a'(n)< m(n)+a(n)-\log3 t_V(n)$
for infinitely many $n$. Here  
$t_V(n)$ stands for the polynomial that upper bounds the running time of 
$V$.  W.l.o.g. we may assume that the running time of  
$\PPNP^A$ protocols does not depend on the oracle.

Then we prove that there is an oracle  $A$ such that this 
protocol does not recognize $L(A)$. To this end we will need the following 
\begin{lemma} \label{l1}
Assume that a procedure $P$ is given that on input (a Boolean matrix $M$ of size $m\times a$,
strings $y,z$) outputs a bit by querying at most $q<2^a/3$ elements of the matrix $M$.
Let $a'$ be a natural number with  
$$	
2^{a'-1}q <(2^{a}/3-q)2^m,
$$  
and $m'$ any natural number. Then there is a heavy Boolean matrix $M$
with  
\begin{equation}\label{eq1}
\prob_{z\in \{0,1\}^{a'}}[\exists y\in \{0,1\}^{m'}, P(M,y,z)=1]\le1/2,
\end{equation}
or a  light Boolean matrix  $M$ with
\begin{equation}\label{eq2}
\prob_{z\in \{0,1\}^{a'}}[\exists y\in \{0,1\}^{m'}, P(M,y,z)=1]> 1/2.
\end{equation}
 \end{lemma}
\begin{proof}
For the sake of a contradiction assume that there is no such matrix. That is, for every heavy $M$
we have~\eqref{eq2}, and for every light $M$ 
we have~\eqref{eq1}.  

Let first  $M$ be all-zero matrix.
We will derive a contradiction by flipping certain $M$'s entries in such a way that
$M$ is still light but there are $2^{a'-1}+1$ pairs $(y,z)$ with pairwise different
first components and with $P(M,y,z)=1$. 
We will find such pairs $(y,z)$ one by one. For each new pair $(y,z)$ we 
will \emph{freeze} all elements of $M$ queried in the computation of
$P(M,y,z)$. This means that we will not change those elements on further steps
and thus the equality   $P(M,y,z)=1$ will remain valid. On each step we will freeze at most $q$ entries and
the assumed inequality 
$$	
2^{a'-1}q <(2^{a}/3-q)2^m
$$  
will guarantee that on each step it is possible to find a row with few frozen entries.
Flipping all  non-frozen elements of that row we get a heavy matrix, which will allow to find a new pair $(y,z)$.

More specifically, we make $2^{a'-1}+1$ steps and on $i$th  step we find one new pair $(y_i,z_i)$.
On the first step we flip all the elements of the  first row of $M$. We obtain a heavy matrix. By assumption the inequality~\eqref{eq2} holds. Hence
there is $z_1$ such that there is $y_1$ with $P(M,y_1,z_1)=1$. 
Fix such $y_1,z_1$. Then freeze all elements of $M$ queried in the computation of
$P(M,y_1,z_1)$. 
Then we again flip all non-frozen elements of the first row and 
$M$ becomes light. 
This follows from the assumed inequality $q<2^{a}/3$, which implies that the number of ones in the first row is less than one third. 

After $i$th step the matrix $M$ is light and at most $iq$ its entries are frozen. Besides that, there are distinct
$z_1,\dots,z_i$ and (not necessarily distinct) $y_1,\dots,y_i$ with
$P(M,y_1,z_1)=\dots=P(M,y_i,z_i)=1$. On 
 $(i+1)$st step we choose a row with minimal number of frozen elements. That number is  less than $2^a/3-q$,
since at most $iq\le2^{a'-1}q <(2^a/3-q)2^m$ entries of the matrix are frozen, that is, on average
less than $2^{a}/3-q$ per row. Then we make all non-frozen elements in that row equal to 1. The resulting matrix is heavy,
as it has more than $2^a-(2^a/3-q)>(2/3)2^a$ ones. By the assumption
we have~\eqref{eq2}. Since $i\le2^{a'-1}$, there is  $z_{i+1}$ that is different from $z_1,\dots,z_i$ and there is $y_{i+1}$ with
$P(M,y_{i+1},z_{i+1})=1$. Pick such $y_{i+1}$ and $z_{i+1}$  and freeze all elements queried in the computation of $P(M,y_{i+1},z_{i+1})$.
Then we make all non-frozen elements of that row equal to 0. The total number of frozen elements in that row is less than  $(2^{a}/3-q)+q=2^{a}/3$,
thus we again get a light matrix.

After  $2^{a'-1}+1$ steps we derive a contradiction: the matrix $M$ is light and yet for more than half of $z\in \{0,1\}^{a'}$ there is $y\in \{0,1\}^{m'}$with
$P(M,y,z)=1$. 
\end{proof}

Let us resume the proof of the theorem. Let us consider the procedure  $P(M,y,z)$ that simulates the run of
$V^A(1^n,y,z)$. If $V$ queries oracle's value on a string   of length $m(n)+a(n)$, then
$P$ queries the corresponding entry of the input matrix $M$.  If $V$ queries oracle's value on a string   of length 
different from $m(n)+a(n)$, then $P$ assumes that the oracle answer is 0.
Note that for all sufficiently large $n$ the running time $t_V(n)$ of $V$ on $1^n$,
and hence the number of oracle queries, is less than $2^{a(n)}/6$. Indeed, $2^{a(n)}$ grows exponentially 
and $t_V(n)$ is a polynomial. For such $n$'s, to meet the assumptions of the lemma, 
it suffices to satisfy the inequality $2^{a'(n)-1}t(n) <2^{a(n)+m(n)}/6$, that is, $a'(n)<a(n)+m(n)-\log3t(n)$.
By the assumption there are infinitely many $n$ satisfying this inequality. Thus there is  
$n$ for which the conclusion of the lemma holds.  
We pick any such $n$ and define the value of the oracle $A$ on strings of length $m(n)+a(n)$ so that the matrix $M=A_n$ satisfy 
the conclusion of the lemma. For all other strings $u$ we let $A(u)=0$.
By construction the chosen  $\PPNP$ protocol  does not recognize $L(A)$. 

Now we have to fool all $\PPNP$ protocols $(m',a',V)$,  
with  $a'(n)< m(n)+a(n)-\log 3t_V(n)$
for infinitely many  $n$. 
This can be done by a standard diagonalization. We enumerate all $\PPNP$ protocols with oracle which  
satisfy  the inequality $a'(n)< m(n)+a(n)-\log 3t_V(n)$
for infinitely many $n$.
We first let $A(u)=0$ for all $u$ and then we perform infinitely many steps. On each step 
we freeze a finite number of oracle values. On $i$th step we fool
$i$th $\PPNP$ protocol  $( m',a',V)$. To this end, we choose   $n$ such that no value of
$A$ on strings of length $m(n)+a(n)$ has been frozen yet,   $t_V(n)<2^{a(n)}/6$ and 
 $a'(n)< m(n)+a(n)-\log 3t'(n)$. Using the lemma, we change oracle values on strings
 of length $m(n)+a(n)$ so that the conclusion of the lemma holds.
Then we freeze all oracle values 
queried by $V$ in the runs on inputs $y,z$ of lengths $m(n),a(n)$, respectively.

Finally, to construct an oracle  $A$ such that the statement of the theorem holds
for all polynomials $m(n),a(n)$, we also enumerate all 
the pairs  $(m,a)$ of polynomials:
 $(m_1,a_1),(m_2,a_2),\dots$ and define 
 $$
 L_{i}(A)=\{ 1^n\mid A_{i,n} \text{ is heavy}\}.
 $$
 Here $A_{i,n}$ is defined exactly as we defined $A_n$ earlier for the pair $(m_i,a_i)$ as $(m,a)$.

Then we enumerate all pairs (a natural number $i$, a $\PPNP$ protocol $(m',a', V)$) and for each such 
pair we make one step. 
On that step we fool the
protocol $(m',a', V)$ as a candidate protocol for $ L_{i}(A)$.  There is one obstacle however. 
In order to place the language $L_i(A)$ in $\MA^A[m_i,a_i]$, we need the property   $P_{m_i,a_i}$ hold for all $i$.
When we change $A_{i,n}$ to fool the protocol $(m',a', V)$ as a candidate protocol for $ L_{i}(A)$, 
we can violate the property $P_{m_j,a_j}$  
for  $n'$ with $m_i(n)+a_i(n)=m_j(n')+a_j(n')$.

To handle this problem, we will split the oracle $A$ into countably many oracles,
one oracle for each pair $(m_i,a_i)$ of polynomials.
 For each 
$i$ we consider the prefix encoding $\hat i$ of the number $i$. It is obtained by doubling each bit
in the binary representation of $i$ and then appending 01 (for instance, $\hat 5=11001101$).
This encoding ensures that for $i\ne j$ strings of the form $\hat i u$ and $\hat j v$ cannot coincide.

Then we change the definition of  the matrix $A_{i,n}$: now it is built from oracle's values 
 on strings of the form $\hat i u$, where the length of $u$ is $m_i(n)+a_i(n)$. 
Changing $A_{i,n}$ does not affect any of the matrices
$A_{j,n'}$ for $j\ne i$. Therefore,  when we change $A_{i,n}$ we cannot violate the  property $P_{m_j,a_j}$  
for any $j\ne i$.
\end{proof}

\section{Refuting Knop's Conjecture under an Oracle}

Recall that Knop's conjecture claims the existence of  a polynomial $p$ such 
for all polynomials $m,a$, if $L\in \heurMA[m,a]$ then $L\in\heurAM[*,p(a)]$.

Notice that if $L\in\heurAM[*,a(n)]$ then there is
a normal $\AM[*,a(2n)]$ protocol that is correct on at least half of inputs of each length
and hence is correct on at least one  input of each length $n$. 
We will build an oracle $A$  under which 
Knop's conjecture is false in a strong way. Under that oracle, for all polynomials $m,a$ there is  
a language in  $\MA[m,a]$ (and hence in $\heurMA[m,a]$) such that every   
$\PPNP[m',a',t']$ protocol 
with $a'(n)<m(n)+a(n)-n-\log_2(6t'(n))$ for infinitely many $n$
errs on \emph{all} inputs of length $n$ for some $n$. 

Let, for instance, $a(n)=n$ and $m(n)=p(n)+n$, where $p$ is an arbitrary polynomial.
We can see that all $\AM^A[*,a'(n)]$ 
protocols for a language in  $L\in\MA^A[m(n),n]$ 
that are correct on at least one input of each length require 
$$
a'(n)\ge m(n)+a(n)-n-O(\log n)=(p(n)+n)+n-n-O(\log n)\gg p(n)= p(a(n))
$$ 
random bits for almost all $n$.  

\begin{theorem}\label{th1a}
There is an oracle $A:\{0,1\}^*\to\{0,1\}$ with the following property.
For all polynomials 
$m(n),a(n)$ 
there is a language $L\in\MA^{A}[m,a]$
such that any $\PPNP^A[m',a',t']$ protocol for $L$
with $a'(n)< m(n)+a(n)-n-\log_2(6t'(n))$ for infinitely many $n$
errs on \emph{all} inputs $x$ of some length $n$. 
\end{theorem}
\begin{proof}[Sketch of proof]
As before, we enumerate all pairs (a polynomial $m$, a polynomial $a$). 
Then we define
$$
L_i(A)=\{x\mid \text{the matrix } A_{i,x} \text{ is heavy}\},
$$
where 
$A_{i,x}$ is a Boolean matrix of size $m_i(n)\times a_i(n)$, $n=|x|$, defined
by
$$
A_{i,x}(u,v)=A(\hat i\hat xuv),\quad |u|=m_i(n),  |v|=a_i(n). 
$$
Note that this time the matrix $A_{i,x}$ depends on the input 
$x$ and not only on its length $n$.
The oracle construction will ensure that  for all $i,x$ the matrix $A_{i,x}$ is either heavy,
or light. Thus for all $i$ the language $L_i(A)$ is in $\MA^A[m_i,a_i]$.

The construction of $ A$ 
will ensure that 
for all $i$ 
and for all $\PPNP$ protocols $(m',a',V)$ with $a'(n)<m_i(n)+a_i(n)-n-\log 6t_V(n) $ 
for infinitely many $n$,
there is  $n$ with the following property:
for all $x$ of length $n$,
\begin{quote}
either $A_{i,x}$ is heavy and 
\begin{align*}
\prob_{z\in\{0,1\}^{a_i(|x|)}}[\exists y\in\{0,1\}^{m_i(|x|)} V^{A}(x,y,z)=1]\le1/2.
\end{align*}
or $A_{i,x}$ is light  and 
\begin{align*}
\prob_{z\in\{0,1\}^{a_i(|x|)}}[\exists y\in\{0,1\}^{m_i(|x|)} V^{A}(x,y,z)=1]>1/2.
\end{align*}
\end{quote}
This implies that the protocol $(m',a',V)$, as a candidate protocol for $L_i(A)$, 
errs on \emph{all} inputs $x$ of a certain length $n$.

The construction of $A$ proceeds in steps where on each step we fool some protocol
  $(m',a',V)$ as a candidate protocol for some language $L_i(A)$. To this end we use the following
\begin{lemma} \label{l3}
Assume that a procedure $P$ is given that on input (a sequence of Boolean matrices $M=(M_1,\dots,M_{2^n})$ of size $m\times a$,
strings $y,z$ and $x\in\{1,\dots, 2^n\}$) outputs a bit by querying at most $q<2^a/3$ elements of the given matrices in total.
Let $a'$ be a natural number with  
$$	
2^{a'}\cdot q\cdot2^n <(2^{a}/3-q)2^m,
$$  
and $m'$ any natural number. Then 
there is a sequence of matrices $M_1,\dots,M_{2^n}$ such that for \emph{all} $x=1,\dots,2^n$
either $M_x$ is heavy
and   
\begin{equation}\label{eq11}
\prob_{z\in \{0,1\}^{a'}}[\exists y\in \{0,1\}^{m'}, P(M,y,z,x)=1]\le1/2,
\end{equation}
or $M_x$ is light and
\begin{equation}\label{eq21}
\prob_{z\in \{0,1\}^{a'}}[\exists y\in \{0,1\}^{m'}, P(M,y,z,x)=1]> 1/2.
\end{equation}
 \end{lemma}
\begin{proof}
The proof is similar to that of Lemma~\ref{l1} and thus we explain only what is the difference.
Now we make $2^n$ stages and on each stage we make $2^{a'-1}+1$ steps.
On stage $x$ we flip only entries of $M_x$ to ensure the requirement for that $x$.
However we need to freeze also entries of other matrices $M_{x'}$ in the case 
$P$ queries their entries in the computation on the input $(M,y_i,z_i,x)$.
Thus the total number of frozen elements, in all the matrices,
can now raise up to  $(2^{a'-1}+1)\cdot q\cdot2^n\le 2^{a'}\cdot q\cdot2^n$.
Since we assume that this number is still less than  $(2^{a}/3-q)2^m$,
on each step each matrix has a row with less than $2^{a}/3-q$ frozen entries.
This we can complete each step.
 \end{proof}
 
 The rest of the proof is similar to that of Theorem~\ref{th1}.
 To fool a protocol
  $(m',a',V)$ as a candidate protocol for the language $L_i(A)$, we apply Lemma~\ref{l3} to the procedure $P$ that simulates
the run of $V^A(x,y,z)$. 
If $V$ queries oracle's value on a string  of the form $\hat i\hat x uv$ where 
$|u|=m_i(n)$ and $|v|=a_i(n)$, then $P$ 
queries $M_x(u,v)$. To fool the protocol $(m',a',V)$, we 
again choose   $n$ such that no value of
$A$ on strings of length $2\log i+2 +2n+2+m_i(n)+a_i(n)$ has been frozen yet,  $t_V(n)<2^{a_i(n)}/6$ and 
 $a'(n)< m_i(n)+a_i(n)-n-\log 6t'(n)$. Using the lemma, we change oracle values 
so that the conclusion of the lemma holds.
Then we freeze all oracle values 
queried by $V$ in the runs on inputs $x,y,z$ of lengths $n,m_i(n),a_i(n)$, respectively.
\end{proof}

\section{Constructing an Oracle under which Transformation of $\MAP$ to $\PPNP$
Protocols Requires Many Random Bits}

Our third result states that 
under some oracle some language in $\MAP[m,a]$ has  
$\PPNP[m',a',t']$ protocols only if  
$a'(n)\ge \max\{m(n),a(n)\}-\log t'(n)-O(1)$.

\begin{theorem}\label{th2}
There is an oracle $A:\{0,1\}^*\to\{0,1\}$ with the following property.
For all polynomials 
$m(n),a(n)$ 
there is a language in $\MAP^{A}[m,a]$ that is outside any 
class $\PPNP^A[m',a',t']$ such that 
$a'(n)< \max\{m(n),a(n)\}-\log t'(n)-O(1)$ for infinitely many $n$. 
\end{theorem}
\begin{proof}
The proof of this theorem is similar to that  of Theorem~\ref{th1}.
However this time heavy matrices  are defined as those for which 
there is an all-one row, and light matrices 
as those in which there are less than one half ones (Fig.~\ref{f2}).
\begin{figure}
\begin{center}
$$
\begin{pmatrix}
0&0&1&0&0&0&0\\
0&0&0&0&0&0&0\\
0&0&0&0&1&0&0\\
0&0&0&0&0&0&0\\
1&1&1&1&1&1&1\\
0&0&0&1&0&0&0
\end{pmatrix} 
\qquad\qquad\qquad\qquad
\begin{pmatrix}
0&1&1&0&1&0&0\\
1&0&1&0&0&1&0\\
0&0&0&0&0&1&1\\
1&0&0&1&0&0&1\\
0&1&0&0&0&0&0\\
0&1&0&0&0&1&1
\end{pmatrix} 
$$
\end{center}
\caption{The matrix on the left is heavy, and  the matrix on the right is light.}\label{f2}
\end{figure}
Instead of Lemma~\ref{l1} we use the following
\begin{lemma}\label{l2} 
Assume that a procedure $P$ is given that on input (a Boolean matrix $M$ of size $m\times a$,
strings $y,z$) outputs a bit by querying at most $q<2^{a-1}$ elements of the matrix $M$.
Let $a'$ be a natural number with  $2^{a'-1}q<2^m$ or $(2^{a'-1}+1)q< 2^{a-1}$
and $m'$ any natural number. Then there is a heavy matrix $M$
satisfying~\eqref{eq1}, or a light matrix $M$  satisfying~\eqref{eq2}. 
 \end{lemma}
\begin{proof}
Assume first that $2^{a'-1}q<2^m$. 
For the sake of contradiction assume that for every heavy matrix $M$ it holds~\eqref{eq2} and
for every light matrix $M$  it holds~\eqref{eq1}.  
Let  $M$ be all-zero matrix.
Flip all bits of the first row. We get a heavy matrix. By assumption we have~\eqref{eq2}.
Choose any  $z_1$ such that there is a $y_1\in \{0,1\}^{m'}$ with $P(M,y_1,z_1)=1$. 
Fix such $y_1,z_1$ and freeze all values of $M$ queried in the run $P(M,y_1,z_1)$.  
Thus we guarantee that  $P(M,y_1,z_1)=1$. 
Then flip again all non-frozen elements of the first row. As $q<2^{a-1}$, now the matrix $M$ is light. 
 
Then we make $2^{a'-1}$ similar steps.
After $i$th step we have at most $iq$ frozen elements of $M$, distinct $z_1,\dots,z_i$ and (not necessarily distinct) $y_1,\dots,y_i$
with $P(M,y_1,z_1)=1,\dots,P(M,y_i,z_i)=1$. 
On  $i+1$st step we choose a  row with no frozen entries 
(such a row does exist, since we assume that $iq\le2^{a'-1}q <2^m$) and flip all elements of that row. We get a heavy matrix
and hence it satisfies~\eqref{eq2}.
Since $i\le 2^{a'-1}$, there is $z_{i+1}$ that is different from  $z_1,\dots,z_i$ and there is $y_{i+1}$
with  $P(M,y_{i+1},z_{i+1})=1$. Freeze all entries of $M$ queried in this computation. Then flip all non-frozen 
elements of the first row. As $q<2^{a-1}$, now the matrix $M$ is light.

After $2^{a'-1}+1$ we get a contradiction as the matrix $M$ is light
and
for more than half of $z\in \{0,1\}^{a'}$ there is $y\in \{0,1\}^{m'}$ with $P(M,y,z)=1$. 

Assume now that $(2^{a'-1}+1)q< 2^{a-1}$. In this case the arguments are simpler. Again,
for the sake of contradiction assume that for every heavy matrix $M$ it holds~\eqref{eq2} and
for every light matrix $M$ it holds~\eqref{eq1}.  
Let  $M$ be all-one matrix and hence it is heavy. By assumption we have~\eqref{eq2}.
Fix  $2^{a'-1}+1$ pairs $(y,z)$ with $P(M,y,z)=1$ and with pair-wise distinct first components. 
Freeze at most $(2^{a'-1}+1)q$ entries of $M$ guaranteeing  
$P(M,y,z)=1$ for all those pairs. Flip all non-frozen elements of $M$. The inequality $(2^{a'-1}+1)q< 2^{a-1}$
implies that less than half entries in each row are frozen. Hence we get a light matrix satisfying~\eqref{eq2}, a contradiction.
\end{proof}

The remaining part of the proof is similar to that of the proof of Theorem~\ref{th1}.
\end{proof}

\section{Open questions}

1. Is it true that for some  polynomial $p$ for all polynomials $m,a$ it holds $\MA[m,a]\subset\AM[*,p(a)]$?

2. Can we prove the inclusion $\MA[m,a]\subset\AM[*,p(a)]$ by algebrizing techniques (for some fixed polynomial $p$)?

3. Can we prove Conjecture 1 by algebrizing techniques?

\section{Acknowledgments}
 
 The author is sincerely grateful to anonymous referees for helpful suggestions.


\begin{thebibliography}{AFWZ95}

\bibitem[AFWZ95]{alon1995derandomized}
Noga Alon, Uriel Feige, Avi Wigderson, and David Zuckerman.
\newblock Derandomized graph products.
\newblock {\em Computational Complexity}, 5(1):60--75, 1995.

\bibitem[AW09]{AW}
Scott Aaronson and Avi Wigderson.
\newblock Algebrization: {A} new barrier in complexity theory.
\newblock {\em {ACM} Trans. Comput. Theory}, 1(1):2:1--2:54, 2009.

\bibitem[Bab85]{babai85}
Laszlo Babai.
\newblock Trading group theory for randomness.
\newblock In {\em 17th STOC}, pages 421--429, 1985.

\bibitem[BGS75]{BGS}
T.~Baker, J.~Gill, and R.~Solovay.
\newblock Relativization of {P}=?{NP} question.
\newblock {\em SIAM J. Computing}, 4(4):431--442, 1975.

\bibitem[BHZ87]{BHZ}
R.~B. Boppana, J.~Håstad, and S.~Zachos.
\newblock Does co-np have short interactive proofs?
\newblock {\em Information Processing Letters}, 25:127--132, 1987.

\bibitem[Kno15]{knop2015}
Alexander Knop.
\newblock Circuit lower bounds for average-case {MA}.
\newblock In Lev~D. Beklemishev and Daniil~V. Musatov, editors, {\em Computer
  Science - Theory and Applications - 10th International Computer Science
  Symposium in Russia, {CSR} 2015, Listvyanka, Russia, July 13-17, 2015,
  Proceedings}, volume 9139 of {\em Lecture Notes in Computer Science}, pages
  283--295. Springer, 2015.

\bibitem[Lau83]{lauteman}
Clemens Lautemann.
\newblock {BPP} and the polynomial hierarchy.
\newblock {\em Inf. Proc. Lett.}, 17(4):215–217, 1983.

\bibitem[Sip83]{sipser-bpp}
Michael Sipser.
\newblock A complexity theoretic approach to randomness.
\newblock In {\em Proceedings of the 15th ACM Symposium on the Theory of
  Computing}, pages 330--335. ACM, New York, 1983.

\bibitem[Ver92]{verPP}
Nikolay Vereshchagin.
\newblock On the power of {PP}.
\newblock In {\em Proc. 7th Annual IEEE Conference on Structure in Complexity
  Theory, Boston, MA}, pages 138--143, 1992.

\end{thebibliography}


\end{document}